\documentclass{epl}

\title{Reentrant spin glass transition in La$_{0.96-y}$Nd$_{y}$K$_{0.04}$MnO$_3$: origin and effects on the colossal magnetoresistivity}

\author{R. Mathieu\inst{1}, P. Svedlindh\inst{1}, \and P. Nordblad\inst{1}}
\institute{
  \inst{1} Department of Materials Science, Uppsala University, Box 534, SE-751 21 Uppsala, Sweden\\
}
\begin{document}

\maketitle

\begin{abstract}
Magnetic, electric and structural properties of La$_{0.96-y}$Nd$_y$K$_{0.04}$MnO$_{3+\delta}$ with
0$\leq y \leq$0.4 have been studied experimentally. A disordered magnetic
state is formed as La is substituted by Nd, reflecting the competition
between
ferromagnetic (FM) double exchange and antiferromagnetic
superexchange interactions. Key structural parameters are identified
and correlated with changes in magnetic and electric properties. By
application of a large magnetic
field, spin disorder scattering is removed, creating a new
magnetoresistance peak at a temperature lower than the first near-$T_c$
peak. Time dependent zero-field-cooled magnetisation measurements have been
performed for $y$=0.4 around this temperature. A reentrant spin glass (RSG)
transition is evidenced, with low field ageing properties in both
the RSG and FM phases, similar to those observed in archetypal spin glass
materials.
\end{abstract}

\section{Introduction}
The doped manganite compounds R$_{1-y}$A$_y$MnO$_3$ (R=rare earth;
A=alkali\-ne earth) have been intensively investigated concerning their
colossal magnetoresistance (CMR) properties \cite{ref00}. Simultaneous
ferromagnetic (FM) and metallic properties in these manganese oxides are
attributed to double exchange \cite{ref0} (DE) interaction between pairs
of Mn$^{3+}$ ($t^{3}_{2g}e^{1}_g$) and Mn$^{4+}$ ($t^{3}_{2g}$) ions.
Substitution of cations with different sizes at the rare earth sites
results in lattice distortions that may influence the ferromagnetic DE and the
antiferromagnetic (AFM) superexchange interactions differently, for
instance increasing the importance of the AFM interactions with
respect to the FM interactions. This can create various forms of magnetic
disorder, thereby also
affecting the transfer integral for itinerant $e_g$ electrons which depends
on the relative spin orientation of the localized $t_{2g}$ moments;
$t_{ij}\propto \cos(\theta_{ij}/2)$, where $\theta_{ij}$ is the angle
between moments $i$ and $j$ \cite{ref1}.
In systems with ferromagnetic DE interaction dominating,
the AFM interactions may still cause magnetic disorder and frustration and
the system is in this case expected to show a low temperature reentrant
spin glass (RSG) phase \cite{ref2,ref5}. When such a system is
cooled from a high temperature, it first exhibits a transition from a
paramagnetic (PM) to a ferromagnetic (FM) phase. If the cooling proceeds, a
transition to a RSG phase occurs. Indications of such a transition
have already been reported for CMR compounds, but without
presenting convincing experimental evidences for its existence. Usually,
the authors consider
that a low temperature cusp in the zero-field-cooled (ZFC) magnetisation,
leading to
a difference between the ZFC and field cooled (FC) magnetisation curves
at low temperatures, is enough to evidence a low temperature
RSG phase \cite{ref1s,ref2s,ref3s,ref4s}. Ac-susceptibility
measurements are sometimes added; a frequency dependent
ac-susceptibility is then attributed to the RSG
phase\cite{ref5s,ref6s,ref7s}.
But, even if these features are indicative of a deviation from a perfectly
ordered ferromagnetic state, they are not sufficient to prove a low
temperature
RSG phase. Moreover, it is possible to envisage other disordered magnetic
phases resulting from a
competition between ferromagnetic DE and antiferromagnetic superexchange
interactions. Low temperature spin canted AF and FM phases were
suggested in the pioneering work of de Gennes \cite{ref1}.
More recent work indicates that the canted phase is unstable against
electronic phase separation and the
formation of hole rich ferromagnetic and hole undoped antiferromagnetic
regions \cite{ref20,ref21}.
Thus, other and complementary experiments must be performed to dissolve the
details of the magnetic phase diagram of the manganese perovskites. For
example in a
disordered and frustrated magnetic system like a spin glass, the
zero-field cooled (ZFC)
magnetisation relaxes in a characteristic way if the
temperature is kept constant;
therefore time-dependent ZFC magnetisation measurements may be used
to study the evolution of the spin configuration and the possible
occurrence of a RSG phase. \\
In the present study, we investigate effects of ion size mismatch
induced by replacing La$^{3+}$ with Nd$^{3+}$ in the manganese oxide
La$_{0.96-y}$Nd$_y$K$_{0.04}$MnO$_{3+\delta}$. A RSG phase transition,
creating a second peak in the
magnetoresistance when applying a sufficiently large magnetic field, is
revealed from ZFC relaxation measurements. The results give evidence in
support of a recent proposal that the AFM superexchange interaction is one
mechanism responsible for spin disorder and that the DE ferromagnetic interaction is more strongly affected by a variation of the $Mn-O-Mn$ bond with cation substitution\cite{ref16}. \\
\section{Samples and Experiments}
Single-phase La$_{0.96-y}$Nd$_y$K$_{0.04}$MnO$_{3+\delta}$, with $y$
ranging from 0 to 0.4 in steps of 0.1 were prepared by solid state
reaction between La$_2$O$_3$, Nd$_2$O$_3$, MnO$_2$ and K$_2$CO$_3$ in
stoichiometric proportions. Substitution of Nd$^{3+}$ at the La$^{3+}$ site,
Nd$^{3+}$ having a smaller ionic radius, modifies the $Mn-O-Mn$ bond angle
and the $Mn-O$ bond length. Structure and composition of the obtained
compounds were checked by room temperature x-ray
diffraction (XRD) on a high resolution STOE transmission diffractometer
equipped with a position-sensitive detector. The scanning range was $10^{o}$-
$90^{o}$ and CuK$\alpha_1$ radiation was used for all data collections.
We observe a change in structure from rhombohedral $R-3C$ to orthorhombic
$Pnma$; Table {\ref{table1}} gives a summary of the structural
data; the change from $R-3C$ to $Pnma$ at an average A-site ionic
radius $<r_A>=1.225$ \AA is in full agreement with earlier
re\-ports\cite{rada}. For $y=0.1$, with $<r_A>=1.224$ \AA, it was
necessary to include a mixture of the rhombohedral and the
orthorhombic structures in the refinements.\\
Low field ac-susceptibility measurements,
$\chi(f,T)=\chi'(f,T)+\chi''(f,T)$,
were performed in a LakeShore ac-susceptometer. Measurements of
resistivity and
magnetoresistance were performed using a Maglab 2000 system from Oxford
Instruments; the magnetoresistance is defined as $(R_0-R_H)/R_0$. Finally,
time dependent ZFC magnetisation
measurements were performed in a non-commercial SQUID magnetometer
\cite{refsquid}. The sample was cooled in zero field
from a temperature above $T_c$ to the measurement temperature. After a
waiting
time ($t_w$), a weak magnetic field ($H$=0.05 to 1 Oe) was applied and the
magnetisation $M(t)$ recorded vs. observation time $t$.
The relaxation rate, defined as $S(t)= H^{-1}$ d$M$/dlog $t$, was derived
from these measurements.
\section{Results and discussion}
\begin{figure}
\onefigure{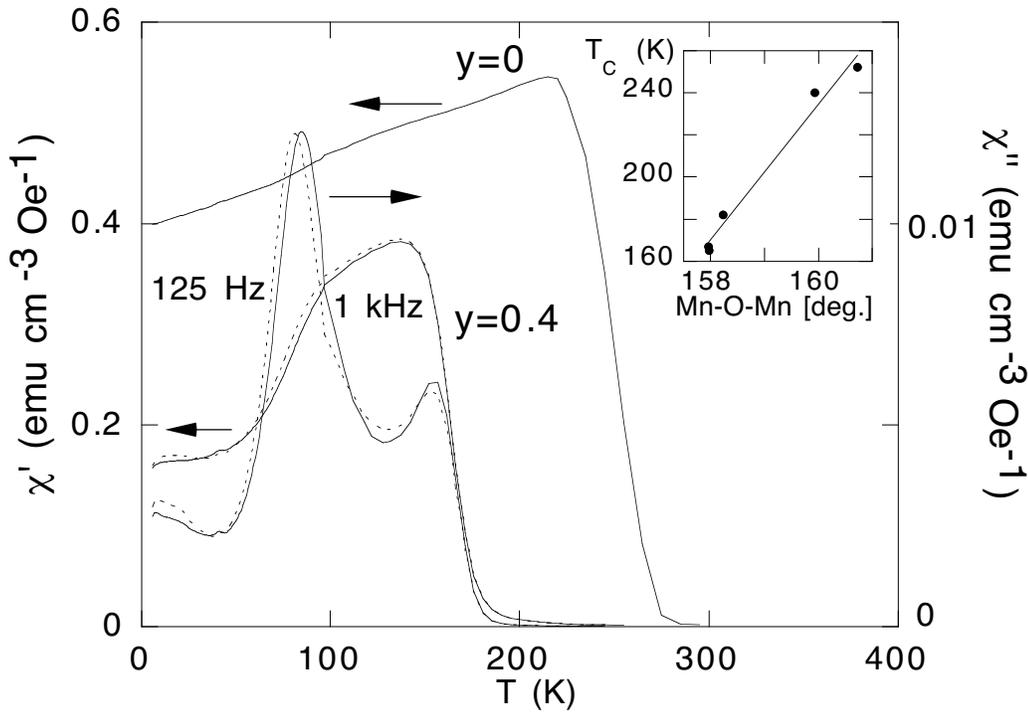}
\caption{In-phase component of the ac-susceptibility for $y$=0 and  $y$=0.4. For $y$=0.4, the out-of-phase component is added, and two frequencies are presented ($f$=125 Hz and 1000 Hz in dotted lines). The insert shows the variation of $T_C$ with the $Mn-O-Mn$ bond angle.}
\label{fig1}
\end{figure}

\begin{figure}
\onefigure{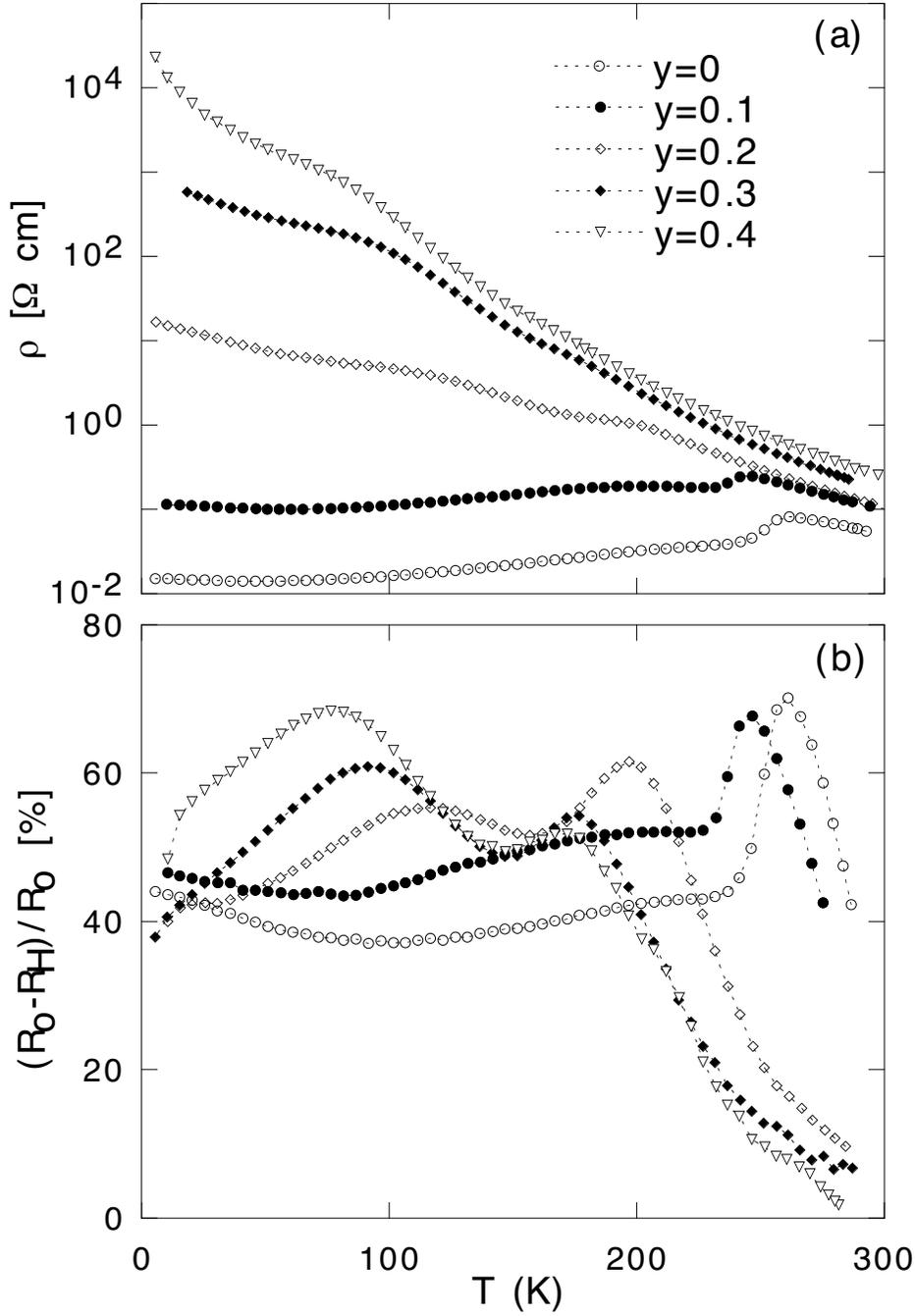}
\caption{Temperature dependence of (a) zero magnetic field resistivity
(log. scale) and (b) magnetoresistance for $H$=5$\times$10$^4$ Oe; the magnetic field is perpendicular to current.}
\label{fig2}
\end{figure}

\begin{figure}
\onefigure{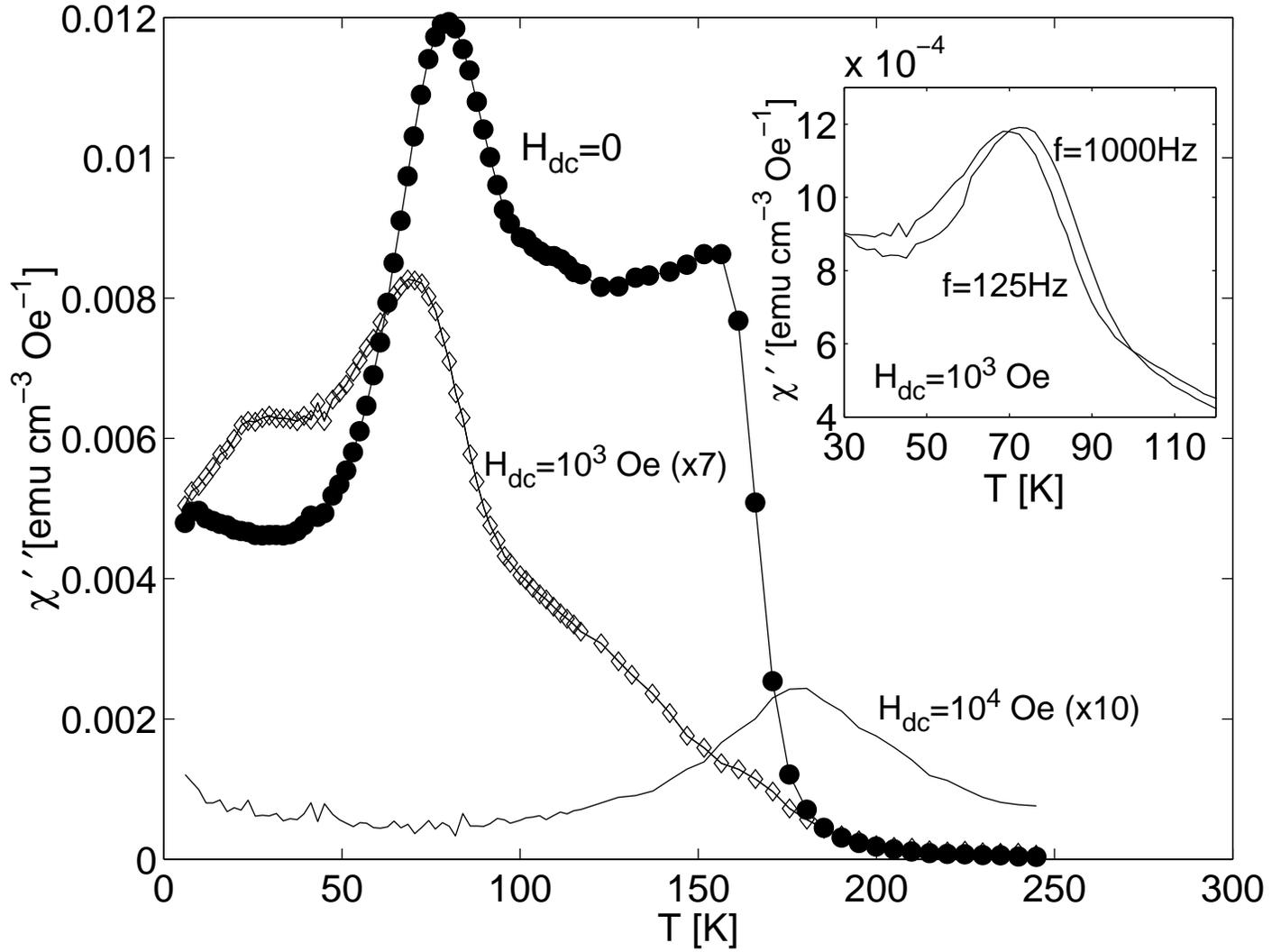}
\caption{Out-of-phase component of the ac-susceptibility for $y$=0.4. $H_{ac}$=1 Oe, and $f$=125 Hz. A dc field $H_{dc}$ is
superimposed. The inset shows the frequency dependence of the low
temperature peak for $H_{ac}$=1 Oe and $H_{dc}$=1000 Oe.}
\label{fig3}
\end{figure}

\begin{figure}
\onefigure{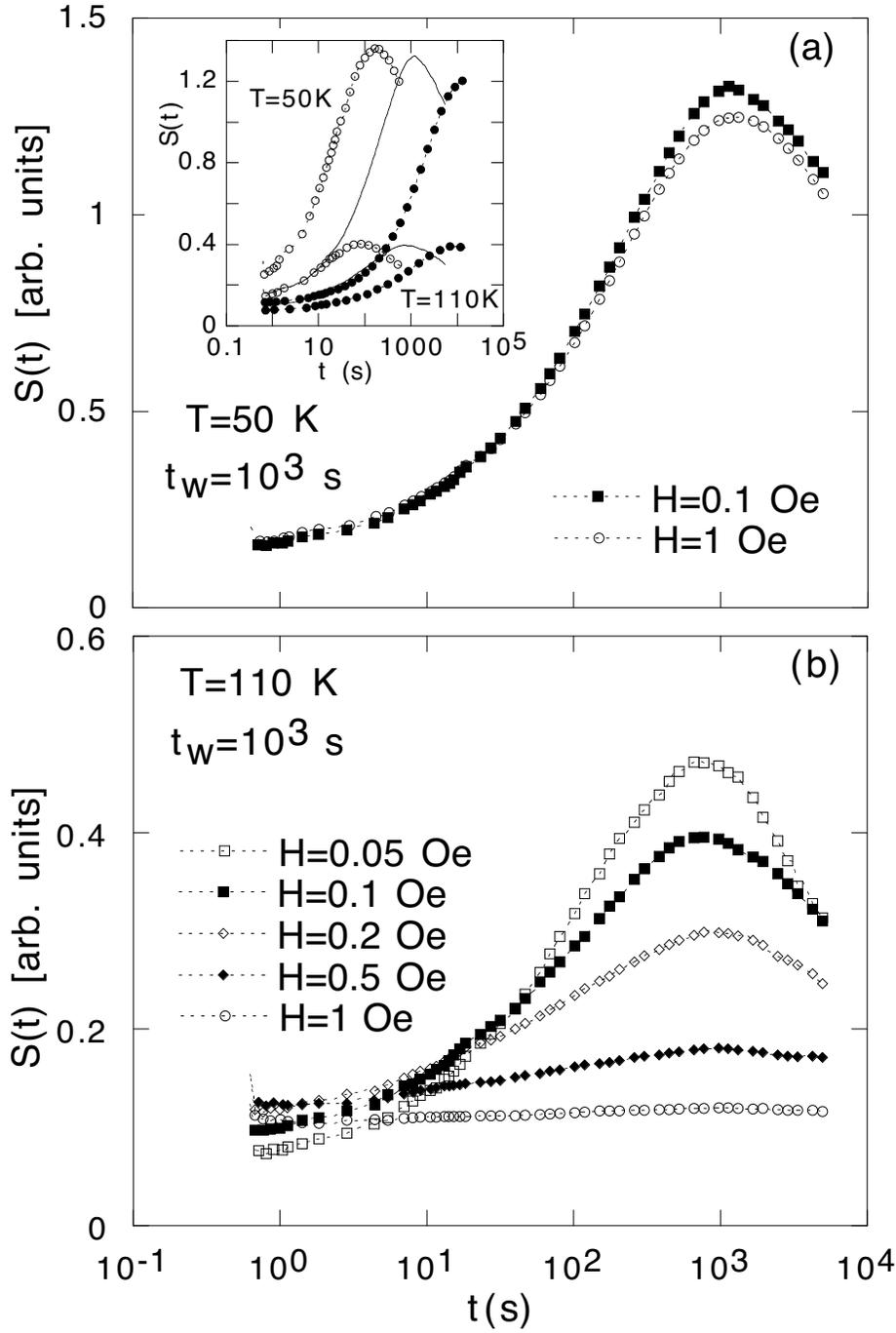}
\caption{Relaxation rate $S(t)$ at $T=50K$ (a) and $T=110K$ (b) for different applied probing fields; $t_w$ = 1000 s. The insert shows the waiting time dependence of $S(t)$ for both temperatures; three different waiting times are used: $t_w$ = 100s (open circles), 1000s (line) and 10000s (filled circles); $H_{dc}$=0.1 Oe.}
\label{fig4}
\end{figure}

The main frame of Fig.~\ref{fig1} shows $\chi'(f)$ and $\chi''(f)$ vs. temperature for the $y$ = 0 and $y$ = 0.4 samples. The Curie temperature ($T_c$), defined from the inflection point in $\chi'(f)$ decreases with increasing Nd doping. At the same time, a frequency dependent knee is created in the in-phase component of the ac-susceptibility at a temperature $T_f(f)<T_c$, corresponding to a large and likewise frequency dependent peak in the out-of-phase component; $T_f(f)$ decreases with increasing Nd doping. This more pronounced frequency dependence of $\chi$ indicates that the magnetic state is more disordered at low temperatures. The magnetic disorder is a consequence of changes in the crystal structure induced by replacing La$^{3+}$ with Nd$^{3+}$. The smaller $Mn-O-Mn$ bond angle (cf. Table {\ref{table1}}) will affect (decrease) not only the DE hopping integral between $e_{g}$(Mn)-2$p_{\sigma}$(O)-$e_{g}$(Mn) orbitals but also the AFM superexchange hopping integral between $t_{2g}$(Mn)-2$p_{\pi}$(O)-$t_{2g}$(Mn) orbitals. However, due to the nature of the dp$\pi$ hybridization, the AFM interaction is expected to be less influenced by changes in bond angle and will thus, relative to the FM DE interaction, increase in importance. The strong effect on the DE interaction is confirmed by the variation of the ferromagnetic transition temperature with bond angle presented in the insert of Fig.~\ref{fig1}.
In passing we note that disorder in the
magnetic interactions may lead to one of several possible, distinctly
different magnetic states; a spin canted state \cite{ref1}, a state with
electronic phase separation between ferromagnetic and
antiferromagnetic regions \cite{ref20,ref21}, or a RSG
state \cite{ref2}. It is important to realize that these possible
magnetic states are expected to display quite different magnetic and
electrical behaviors, something which will discussed more in detail below.
Fig.~\ref{fig2} (a) presents the zero-field resistivity vs. temperature
for all samples. A metal-insulator
(M-I) transition occurs for Nd dopings $y=0$ and  $y=0.1$ at a temperature
close to $T_c$. As the Nd doping
further increases the resistivity increases and no clear M-I transition is
observed; the resistivity at low temperatures increases with decreasing
temperature. By application of a large magnetic field, the spin disorder
scattering induced by competing FM and
AFM interactions is removed, creating a new magnetoresistance peak at
a temperature lower than the first near-$T_c$ peak (see Fig.~\ref{fig2}
(b)); for $y$=0.4, this second peak is larger in magnitude than the first
one. The second peak in the magnetoresistance appears at a temperature
close
to $T_f(f)$, suggesting a correlation between these different experimental
observations. This is confirmed re-measuring $\chi''$ with a superimposed dc field, as shown in Fig.~\ref{fig3}. The dissipation is strongly suppressed by the dc
field; at $H_{dc}$=$10^3$ Oe, the frequency dependent (cf. insert of Fig.~\ref{fig3}) peaks in $\chi''$
are still observed, shifted slightly to lower temperature, while at
$H_{dc}$=$10^4$ Oe, the dissipation is very weak and the RSG like features
are no longer observed. This indicates that, since the second peak in the
magnetoresistance at $T_{f}$ continues
to develop increasing the field further (cf. Fig.~\ref{fig2} (b)), weak
magnetic disorder remains,
in the sense that the spin system is not fully collinear even after
suppressing the slow spin dynamics of the RSG phase. \\
The frequency dependence of the ac-susceptibility and the absence of a M-I
transition for larger
Nd dopings suggest that the ferromagnetic spin arrangement and
the electronic configuration are perturbed. A transition from a
FM metallic to an AFM partially charge ordered state with electronic phase
separation of antiferro- and ferromagnetically ordered regions will result
in a sharp
decrease of $\chi'$ at the transition temperature as well as a
likewise sharp increase of the resistivity\cite{reftok1}. Moreover,
in partially
charge ordered systems,
application of a magnetic field drastically modifies the transport
behavior; a relatively modest magnetic field will turn the insulating state
to a metallic state\cite{reftok1,reftok2}. These characteristics are not
observed in the presently investigated manganite system.\\
Another explanation for the observed spin-disorder is that a spin glass
phase appears at low temperatures. One actually refers to a reentrant spin
glass phase
since the disordered magnetic phase appears at temperatures below where an
ordered (ferromagnetic) phase first appears. This kind of samples have
been extensively studied in the past, and it has been shown that
RSG samples exhibit similar relaxation properties as conventional spin
glasses\cite{ref6}. In particular, the time dependence of the ZFC
magnetisation depends on the time $t_w$ used to equilibrate the spin system
before applying a magnetic field.
This dependence is revealed by an inflection point in the $M(t)$ vs. log
($t$) curve, corresponding to a maximum in
the relaxation rate $S(t)$ curve, at an observation time close
to the waiting time $t \approx t_w$. The phenomenon is called ageing, and
is an inherent property of disordered and frustrated spin systems, while
this property is not expected to be part of the spin canted phase
originally described by de Gennes \cite{ref1}.\\
Time dependent ZFC magnetisation measurements have been performed
for the sample $y$=0.4 to study the behavior of the magnetic relaxation at
temperatures
both below and above the temperature where the disordered magnetic
state appears. Fig.~\ref{fig4} (a) shows the relaxation
rate $S(t)$ for $T=50$K (below $T_f(f)$)
and $T=110$K (above $T_f(f)$), for different magnetic fields. The sample
displays clear ageing behavior
at both temperatures, with a waiting time dependence similar to that
observed for ordinary spin
glasses \cite{ref6} as shown in the insert of Fig.~\ref{fig4}, explicitly
proving the RSG nature of the low temperature transition. The
magnitude of the observed ageing phenomenon as well as the magnitude of
the peak in $\chi''(T)$ prove the effect to be
intrinsic to the material rather than confined to a possible spin disorder
at grain boundaries. Also, as shown in
Fig.~\ref{fig4} (b) the relaxation rate for $T=50$K indicates linear
response of the magnetisation up to
(relatively) high fields in the RSG phase. In the ferromagnetic phase (see
Fig.~\ref{fig4} (b)), the response is non-linear even in weak magnetic
fields and the ageing phenomenon is gradually
suppressed  when increasing the magnetic field, as it has also been
observed in conventional
reentrant spin glass materials \cite{ref7}.

\section{Conclusion}
The resistivity of the La$_{0.96-y}$Nd$_y$K$_{0.04}$MnO$_{3+\delta}$
system increases with increasing $y$
while the Curie temperature decreases, both observations indicating a
lesser degree of
FM double exchange interaction in favor of AFM superexchange interaction.
The competition
between FM and AFM exchange interactions results in spin disorder and
frequency dependent
characteristics of the ac-susceptibility; a frequency dependent peak in
$\chi''(f,T)$ is observed
at a temperature $T_f(f)<T_c$. The spin disorder is also evidenced by a
second
magnetoresistance peak at a temperature close to $T_f(f)$. Time dependent ZFC
magnetisation
measurements show ageing phenomena, both below and above  $T_f(f)$, revealing
the identity of
the low temperature disordered magnetic state to be a reentrant spin glass
phase. These results show the significance of the AFM
superexchange interaction between localized $t_{2g}$(Mn) moments and how by
substitution of cations it is possible to control its magnitude
relative to the magnitude of the ferromagnetic DE interaction.

\acknowledgments

Financial support from the Swedish Natural Science Research Council (NFR) is acknowledged. The authors are grateful to Y. Guo for sample preparation and to Prof. Y. Anderson for help with structural analysis.

\pagebreak

\begin{table}
\caption{Structural data for all compositions: average A-site ionic
radius $<r_A>$ (\AA), average bond length $Mn-O$ (\AA), bond angle
$Mn-O-Mn$ (degrees), and Curie temperature $T_c$ (K); $<r_A>$ is
calculated using data from \cite{sha}.}
\label{table1}
\begin{center}
\begin{tabular}{cccccc}
$y$&Struct.&$<r_A>$&$<d_{Mn-O}>$&$\theta_{Mn-O-Mn}$&$T_c$\\
0&R-3C&1.229&2.080&160.73&252\\
0.1&(R-3C,Pnma)&1.224&(1.970,1.973)&(159.75,160.11)&240\\
0.2&Pnma&1.218&1.977&158.23&182\\
0.3&Pnma&1.213&1.976&157.98&165\\
0.4&Pnma&1.208&1.972&157.97&167\\
\end{tabular}
\end{center}
\end{table}

\end{document}